# νAncestor: an artificial satellite-borne star for accurate frequency calibration of ground-based EPRV spectrographs

F. Pepe[a], M. Bugatti[a], A. Baur[b], J. Berney[c], E. Bozzo[a], C. Broeg[d,e], B. Crazzolara[f], A. Deep[g], F. Droz[c], J. Figuereido[b], H. Haile[a], L. Jolissaint[b], S. Lecomte[c], J. Moerschell[h], Ch. Mordasini[d], E. Obrzud[c], C. Praplan[h], M. Sarajlic[d,e], B. Soja[f]; M. Van den Broeck[i]

[a]Département d'astronomie de l'Université de Genève, Chemin Pegasi 51, 1290 Versoix, Switzerland; [b]Industrial Automation Institute, HES-SO, Cheseaux 1, 1400 Yverdon-les-Bains, Switzerland; [c]Swiss Center for Electronics and Microtechnology (CSEM), Rue de l'Observatoire 58, 2000 Neuchâtel, Switzerland; [d]Space Research and Planetary Sciences, Physics Institute, University of Bern, Sidlerstrasse 5, 3012 Bern, Switzerland; [e]Center for Space and Habitability, University of Bern, Gesellschaftsstrasse 6, 3012 Bern, Switzerland; [f]Institute of Geodesy and Photogrammetry, ETH Zurich, Robert-Gnehm-Weg 15, 8093 Zurich, Switzerland; [g]ESA/ESTEC, Keplerlaan 1, 2201 AZ Noordwijk, Netherlands; [h]Systems Engineering Institute, HES-SO, Industrie 23, 1950 Sion, Switzerland; [i]ESOC Mission Analysis Section - Telespazio Germany under GOF9 Contract for Deliverables and Industrial Support to ESOC.

## ABSTRACT

The accuracy of state-of-the-art Extreme Precision Radial Velocity (EPRV) spectrographs depends on the access to extremely precise and stable wavelength calibration sources. There are several available calibration sources (e.g., emission lamps, laser frequency combs, reference cavities) that can be used to calibrate an astronomical spectrograph. However, the calibration as it is currently performed is always 'local'. In the proposed talk we will present the νANCESTOR concept that proposes an accurate (absolute) and common wavelength calibration for astronomical high-resolution, high-precision spectrographs by embarking an optical frequency comb on-board a satellite equipped with an actively pointing telescope and precision orbitography. This calibration satellite shall be available and serve EPRV spectrographs in all major observatories around the world.

**Keywords:** EPRV, high-fidelity spectrographs, absolute wavelength calibration, calibration satellite

## 1 INTRODUCTION

Study of exoplanets, searches for the Earth analogues and investigation of the nature of the Universe are the main subjects of the high-fidelity astronomical spectroscopy. Since the discovery of the first exoplanet orbiting a solar-type star by Michel Mayor and Didier Queloz [1] using the radial velocity technique, the field has been thriving, with new, often unexpected, discoveries emerging one after the other. High-fidelity spectroscopy has become the main technique to detect planetary atmospheres and bio-signatures, to characterise host stars, to measure the expansion rate of the Universe, or even to test fundamental physics by measuring the possible variability of fundamental constants. Presently, there are several tens of high-resolution astronomical spectrographs scattered all around the world [2]. Their scope ranges over various domains of astrophysics, from the search and characterization of habitable exoplanets, through fundamental physics, e.g., the study of the variability of physical constants, to the direct measurement of the expansion of the Universe. These fundamental questions of astrophysics can all be addressed by high-fidelity spectroscopy provided that extreme accuracy is achieved.

Scrupulous investigation of the constituents of the Universe requires sophisticated tools and scientific methods. One of the most powerful tools to study the Universe is astronomical spectroscopy which relies on decomposing the light according to its wavelength to create optical spectra. Studying spectral features – emission and absorption lines – gives us the knowledge of the physical properties of the observed celestial objects. It can reveal existence of other worlds otherwise hidden from a direct view of telescopes, can probe the history of the Universe and can help to reveal the nature of dark matter. All these thrilling subjects of study require however dedicated instruments able to provide precision, and often accuracy.

A consortium of Swiss institutions has been granted support to carry out, within Switzerland, a feasibility study for a free-flyer experiment, νANCESTOR, which will bring for the first time in space a laser frequency comb (LFC) aiming at the calibration of EPRV spectrographs. The goal of this mission is to provide a universal calibration source to all ground spectrographs dedicated to high resolution radial velocity (RV) measurements. The ultimate science objective is to drastically improve the precision achievable with the RV technique down to a few cm/s on a decade-long timescale, giving the unprecedented possibility to the interested community to eventually combine, on a wide timespan, data acquired with different spectrographs. The Swiss consortium involved in the νANCESTOR study includes the University of Geneva

(UoG), the University of Bern (UBE), the Institute of Geodesy and Photogrammetry (at ETH Zurich), the Haute Ecole Spécialisée de Suisse occidentale (HSE-SO), and the Swiss Center for Electronics and Microtechnology (CSEM).

## 2 SCIENCE CASES

### 2.1 Exoplanet detection

The exploration of the Universe for the existence of exoplanets – planetary companions to stars outside of our Solar System – is currently in its golden age. As of July 2024[1], there are over 5600 confirmed exoplanets, within over 4000 planetary system [3]. The field has provided us with discoveries no one could have imagined – existence of worlds that were once thought to be physically impossible to exist (e.g. [4]). Discoveries that challenged not only our understanding of the Universe and its constituents, but also the understanding of our own backyard – the Solar System, and our home – Earth. A great part of these discoveries can be attributed to astronomical spectroscopy, more specifically the radial velocity method.

The first confirmed detection of an exoplanet orbiting a solar-type star was performed using the radial velocity method [1]. Detection of planetary companions relies on a measurement of periodic shifts in the spectra of the host star – shifts that are due to the reflex motion of the star imposed by the presence of a planet. The smaller the planet, and the further away it is orbiting from its host star, the smaller the spectral shift in the observed spectra and the more difficult it becomes to detect such a planet. Since the very beginnings of the development of this technique in the middle of the last century, the precision of the measurements improved from several hundreds of m/s to < 0.5 m/s in the optical domain [5][5], a level where detection of rocky worlds is possible. One of the main drivers that fuels this progress is a possible existence of an Earth analogue – an Earth-sized planet orbiting a solar-type star at a similar distance as Earth. The detection of such a planet requires a measurement repeatability in terms of radial velocities of < 0.1 m/s, or $3 \times 10^{-10}$ in relative frequency shift (equivalent to radial-velocity change of less than 0.1 m/s) over time scales of years!

Recently, we have been able to confirm with the ESPRESSO spectrograph a known planet around our closest neighbour, Proxima Centauri, and to detect a new planet of only 0.25 Earth-masses on a 5-day orbit [7] (Figure 1). These observations showed a measurement precision of less than 30 cm/s, which is remarkable. However, it must be noted that it required a high sampling, a lot of telescope time, and the use of one single instrument. This precision could only be guaranteed because the orbital period, and thus the time span of the observations, is short (several weeks). The detection of an Earth analogue on a one-year orbit would require a 10 times better repeatability (3 cm/s) over years.

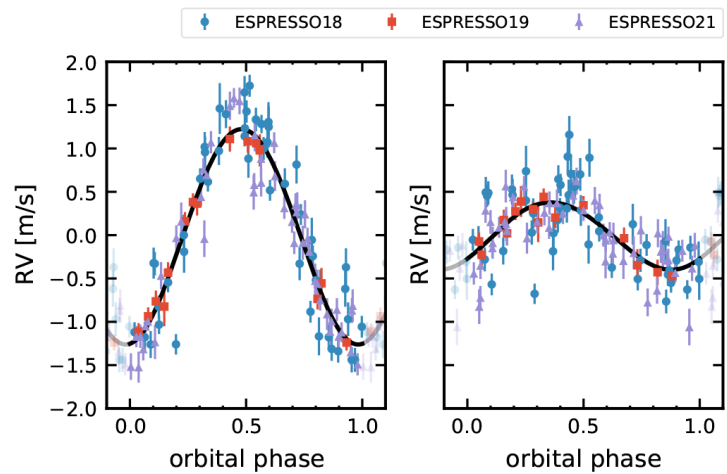

*Figure 1: Radial velocity (RV) data and best-fit keplerian model from ESPRESSO, phase-folded to the orbital periods of Proxima b (left panel) and Proxima d (right panel). From Faria et al. 2022[7]*

### 2.2 Exoplanet characterisation

The radial velocity method is an invaluable tool not only for exoplanet detection but also for the characterisation. It remains the only method that is able to provide constraints on the mass of the exoplanet with the possible exception of resonant systems that allow mass determination through transit timing. The left panel of the Figure 2 shows the mass-radius diagram of small exoplanets as of August 2022. Only planets published in a refereed journal with a mass precision better than 25% and a radius precision better than 8% are shown. The plot highlights the contributions from HARPS, HARPS-N and ESPRESSO spectrographs to our current knowledge of the small planet population. Clearly, these three instruments are leading the worldwide effort in obtaining precise mass measurements. The right panel of the Figure 2 presents an overview of the number of measured exoplanet masses per instrument. Here instruments were assigned to individual planets based on their actual contributions to the mass measurements as described in the most recent publications. HARPS, HARPS-N and ESPRESSO together have provided the majority of all measurements so far, and this share will likely increase further in the near future as the ESPRESSO GTO program has several multi-planet systems almost ready for publication. The impact of ESPRESSO will be particularly important in the 1 – 5 Earth mass regime thanks to its superior instrumental precision and photon collecting power. The properties of the super-Earth population, e.g. the transition between the rocky to gas-rich planets, and the effects of stellar mass and irradiation can thus be studied in detail in the coming years. It becomes evident, however, that the exploration of the lower end of the planet mass range requires improved precision and accuracy.

---

[1] https://exoplanetarchive.ipac.caltech.edu/docs/counts_detail.html

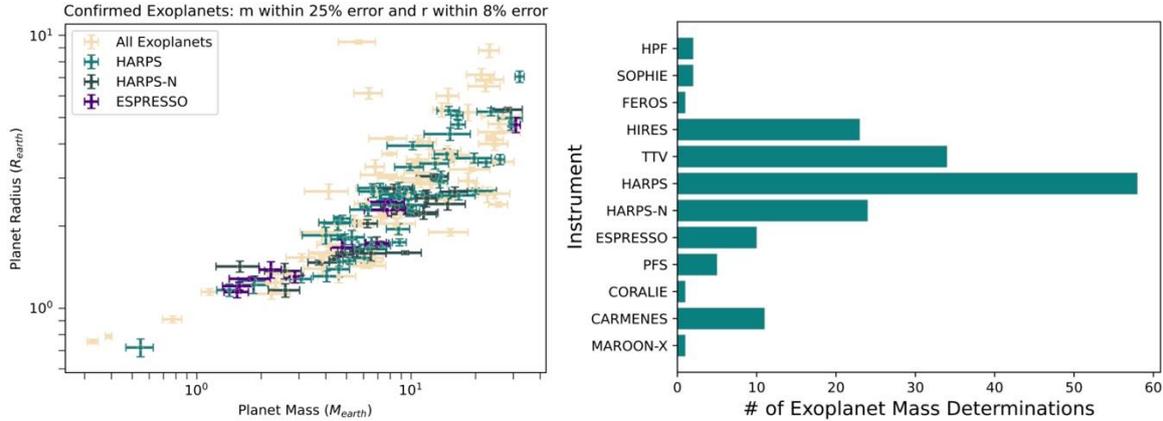

*Figure 2: (Left) Mass-radius diagram of small exoplanets as of August 2022. Only planets published in a refereed journal with a mass precision better than 25% and a radius precision better than 8% are shown. (Right) Number of measured planet masses per instrument. Here instruments were assigned to individual planets based on their actual contributions to the mass measurements as described in the most recent publications. Both figures highlight the contributions from HARPS, HARPS-N and ESPRESSO to our current knowledge of the small-planet population.*

In addition to the information on the planetary mass, measurements of the spectrum of an exoplanet give direct information on the planet's chemical composition and conditions governing its atmosphere. This in turn gives an insight into the planet formation, evolution, and habitability [8].

Using HARPS we (A. Wyttenbach, C. Lovis, F. Pepe) obtained for the first time a clear sodium detection in the upper atmosphere of HD189733 b [9]. The high spectral resolution and quality of the data allowed us to resolve the sodium absorption feature and model it with an appropriate transmission code, which yields the temperature profile in the upper layers of the atmosphere. As a result, a positive temperature gradient with altitude extending towards the thermosphere was derived. A second sodium detection was obtained on the hot Saturn WASP-49 b [10]. In this case the sodium signature was found to be much more extended (up to 1.5 planetary radius), hinting at a possible evaporation of the upper atmosphere due to the strong stellar irradiation.

During the past years, we have been extending our methodology to transiting exoplanets down to the mass of Neptune using HARPS, HARPS-N, and ESPRESSO. Dedicated observing programmes like HEARTS@HARPS and SPADES@HARPS-N (PI: D. Ehrenreich, Co-PIs: F. Pepe, C. Lovis) were awarded time on both instruments and led to numerous detections. On ESPRESSO, we decided to devote one third of the GTO time reserved for exoplanet research to the characterization of planetary atmospheres through transit spectroscopy. This latter programme has been extremely successful and set new standards thanks to the photon collecting power and instrumental stability of ESPRESSO. We could perform time-resolved transmission spectroscopy for the first time and reported the possible detection of iron condensation ('rain') in the ultra-hot atmosphere of WASP-76 b [11] (Figure 3). Among other relevant results obtained with ESPRESSO and its atmospheric working group (D. Ehrenreich, V. Bourrier, F. Pepe, C. Lovis), we can further mention our latest publication that reports on the detection of barium in the atmospheres of the ultra-hot gas giants WASP-76b and WASP-121b, and of Co and Sr+ on WASP-121b [12]. Clearly, the full potential of ESPRESSO for transit spectroscopy has not been exploited yet. Neptune-mass planets are now within reach. Many TESS candidates remain to be explored with ESPRESSO, and the advent of the PLATO mission promises a strong demand for ESPRESSO-like instruments to study planetary atmospheres in the near future.

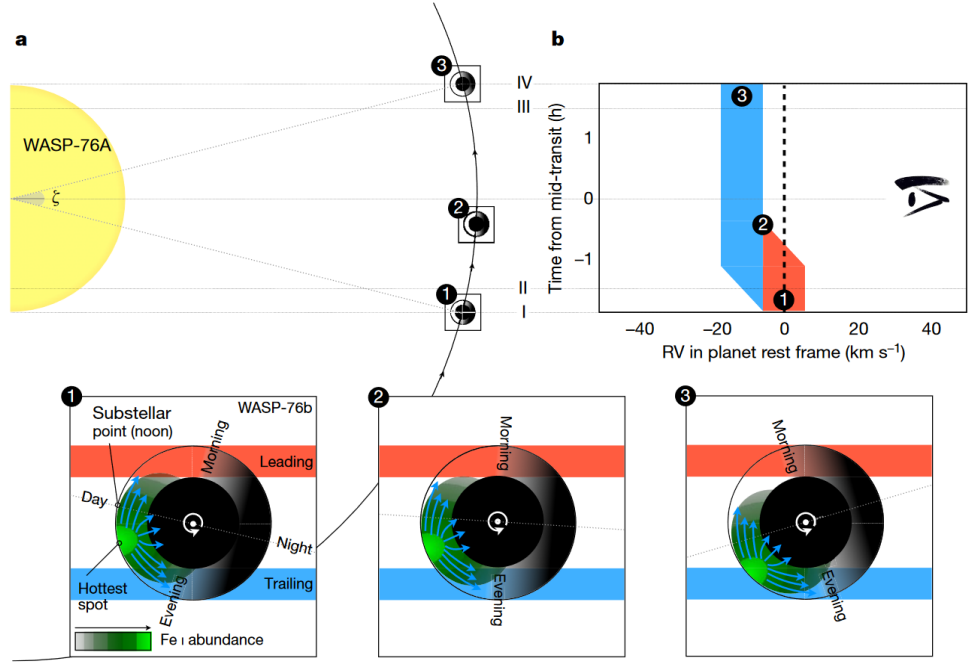

*Figure 3: Polar view of the WASP-76 system. a, The star WASP-76 and planet WASP-76b are represented to scale in size and distance. The planet is shown at different transit stages, with the transit contacts I, II, III and IV. During transit, the angle ζ between the planet terminator and the line of sight (dashed line in the middle) changes by 2 arcsin R⋆/a = 29.4°, where a is the semi-major axis. b, Sketch of the absorption signature observed during transit, in the planet rest frame. The numbers refer to the insets. (1) During ingress, iron on the dayside is visible through the leading limb and creates an absorption around 0 km/s. The trailing limb enters the stellar disc and progressively blueshifts the signal. (2) The signal around 0 km/s disappears as soon as no more iron is visible in the leading limb. Only the trailing limb contributes to the signal, which remain blueshifted around −11 km/s. (3) The signal remains at this blueshifted velocity until the end of the transit (from Ehrenreich et al. 2020).*

## 2.3 Cosmology: the Sandage–Landau test

Even though the expansion of the Universe was discovered almost 100 years ago, the argument over the rate at which the universe is expanding is not over. Despite development of many techniques for non-direct measurement of the expansion rate of the Universe [13]-[16], the existence of substantial discrepancies between the results suggests that there is something missing in our understanding of the Universe. A direct, model-independent measurement could shed light upon this problem. This is where astronomical spectroscopy steps in: as the accelerating expansion of the Universe implies that the relative velocity between two arbitrary points increases with time, the redshift of the spectral features of an observed distant light source will slightly change with time. A change that is so small that thus far a direct measurement of the expansion of the Universe was elusive. It has been theoretically shown that in order to measure the expansion rate of the Universe, a measurement precision of a few cm/s over decades is required [17].

## 2.4 Constituents of the Universe: test of the invariance of fundamental physical constants

Fundamental couplings determine the structure of atoms – the building blocks of matter. Tests of the stability of such fundamental couplings are the most direct way of testing the Einstein Equivalence Principle and constrain dynamical dark energy models. Fundamental couplings such as the fine structure constant α, which determines the strength of the electromagnetic interaction, and the proton-to-electron mass µ, which is a measure of the interplay of the quantum chromodynamics and electroweak physics, are at the forefront of the temporal variation measurements. Observations of high-resolution spectra of gas clouds in the line of sight of bright quasars can be used to derive the fine-structure constant α and the proton-to-electron mass ratio µ [18]-[20]. As light from distant quasars travels through interstellar medium, it is absorbed at different distances by gas clouds and imprints atomic absorption lines into the spectra. A comparison of these tiny line shifts from different absorption lines that have distinct sensitivity to α and µ gives information about the change in the physical constant values at different cosmological distances. These measurements require extremely accurate spectrographs and hence appropriate calibration sources.

# 3 RATIONALE FOR NUANCESTOR

High-fidelity spectroscopy in general, and EPRV Spectrographs in particular, hinge on access to highly precise and stable wavelength calibration sources. Several calibration sources, such as emission lamps, laser frequency combs, and reference cavities, can be used for this purpose. However, current calibration methods have limitations:

1. **Precision vs. accuracy**: most astronomical spectrographs achieve high precision in wavelength calibration. While relative precision suffices for some exoplanet studies, detecting Earth analogues would require long-term repeatability and possibly the sharing and comparison of data taken with different spectrographs. Therefore, absolute calibration is crucial. In fact, long-term repeatability is crucial.

2. **Lack of a common reference**: different spectrographs use different calibration references, making it impossible to directly compare or combine data from multiple instruments. This limitation hinders long-term measurements and reduces overall measurement sensitivity.

3. **Partial optical path calibration**: traditional calibration methods only account for part of the optical path from the telescope to the spectrograph. Errors introduced before the calibration point (e.g., by the atmosphere or telescope front-end) are not detected, potentially affecting data quality.

The νANCESTOR project aims at addressing these issues by deploying an optical frequency comb onboard a satellite to provide a few optical frequency anchors using an absolute laser source operating in multiple wavelength bands. By maintaining a universal reference, the satellite would improve the precision and accuracy of exoplanet studies and other astronomical research, ultimately benefiting all researchers by fostering greater collaboration and data sharing. In fact, astronomical spectrographs worldwide could access this light source for cross-comparison with local calibration sources and standardize calibration across different instruments, allowing for data comparison and merging. Despite the satellite's role, each spectrograph must maintain an on-site calibration source to account for any discrepancies or issues that may arise locally, and the satellite could act as the calibrator for the on-site calibration sources.

## 3.1 Service to the community

We created a non-exhaustive list of astronomical observatories around the world with high-fidelity and EPRV spectrographs that could benefit from accessing a space-based, absolute wavelength-calibration source. Figure 4 and Figure 5 show the EPRV spectrographs on a world map and their wavelength range coverage.

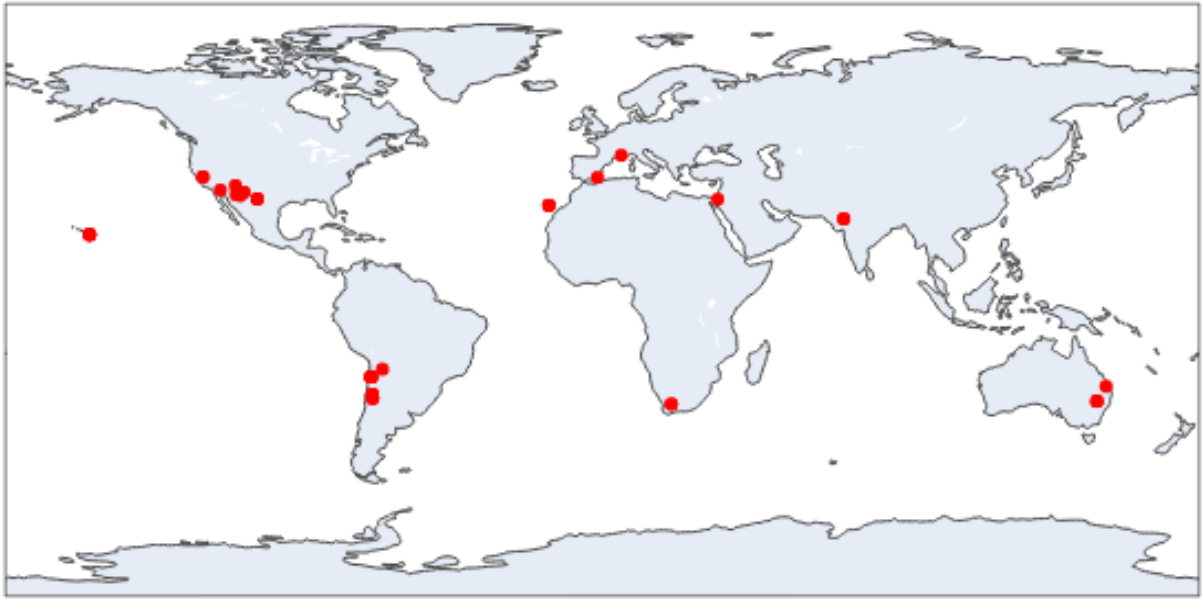

*Figure 4: World map of major astronomical observatories indicated by a red dot*

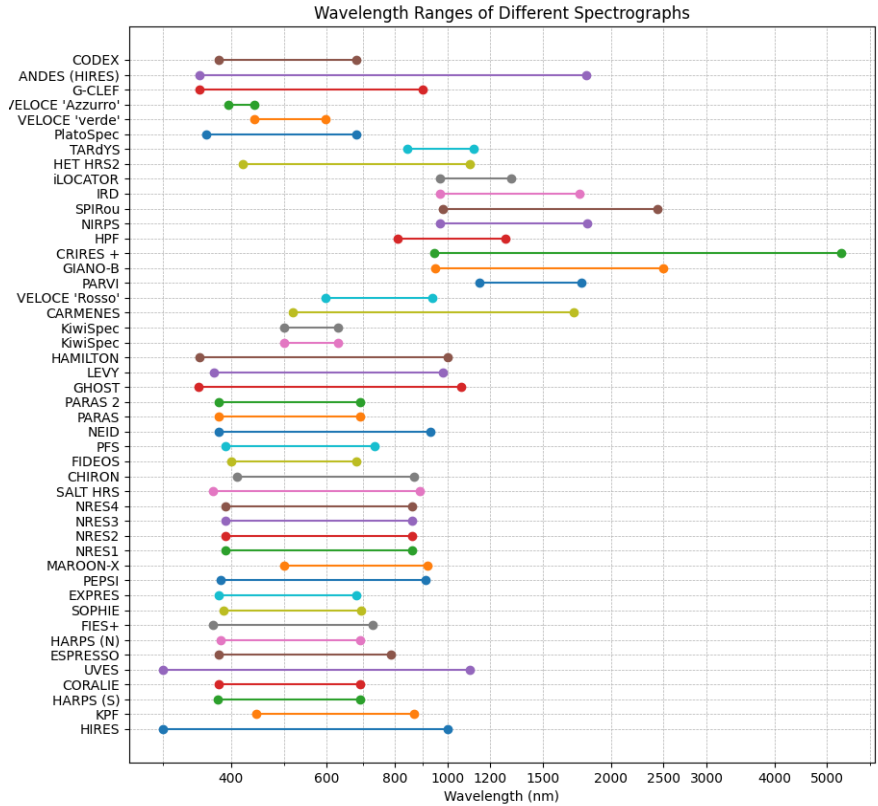

*Figure 5: List of high-fidelity and EPRV spectrographs and their respective wavelength coverage.*

## 4 MISSION CONCEPT

While a detailed science requirement document is in preparation as part of the on-going project study phase (see §6), we have formulated a few initial considerations, or top-level scientific requirements, to set the frame of the future satellite mission:

- The primary objective is to serve EPRV spectrographs in the near-infrared (NIR) and visible (VIS) spectral ranges, specifically from 520 nm to approximately 1550 nm. As a goal, the spectral band shall be extended in the blue down to approximately 380 nm.
- The satellite should be accessible from any Observatory at least once every second night, weather permitting.
- The satellite shall be accessible by an individual instrument and its telescope for at least 15 minutes (continuous) during each contact.
- The space-based calibration sources shall be able to produce an SNR > 1'000 per spectral bin (assuming a resolving power of R = 100'000).

During the contact with the νANCESTOR spacecraft, the ground astronomical spectrographs will have access to a universal light source for cross-comparison of the local calibration source, ensuring that also the full telescope–spectrograph optical path is fully calibrated and any uncertainty introduced in the instrumentation chain is identified and accounted for. The mission, including the payload and the platform, is being designed to achieve an ultimate required RV precision for all ground spectrographs of 10 cm/s (with a goal set at 1 cm/s). More details are provided in the sections below.

### 4.1 Payload overview

The νANCESTOR payload comprises four main sub-systems, namely: (1) a high repetition rate optical frequency comb generating the light that is broadcasted by the satellite; (2) a low repetition rate frequency comb that is referenced to the signal system time provided by a GNSS and serves as reference for the broadcasted frequency comb; (3) a GNSS receiver, providing the precise orbit determination and the disciplined frequency reference to the low-frequency comb; (4) an on-board telescope with a dedicated pointing system to achieve on the ground an adequate footprint for the laser frequency comb signal.

## 4.2 Laser Frequency Comb (LFC)

Based on the Swiss consortium expertise in astrocombs [21]-[23] and space instrumentation involving lasers [24], we selected an architecture based on the so-called electro-optic modulation comb approach and nonlinear frequency conversion to generate different spectral bands. This architecture is under detailed study and so far considered as the most promising at this stage in view of space qualification and general performances.

The laser system will be based on telecom-grade technology that has already undergone two decades of research and development and at least partially has already been tested for space-based applications. The core of the system is a continuous-wave (CW) laser which is electro-optically modulated (EOM). EOM generates a multitude of sidebands in the frequency domain, essentially creating a narrowband laser frequency comb. Following, to access various bands that are useful to many different astronomical spectrographs, nonlinear-frequency conversion will be used. The advantages of the EOM for the use in astronomical spectrograph calibration are the following:

- Generation of optical spectra composed of equidistant laser lines with line separation directly at high frequencies resolvable by astronomical spectrographs and referenced to atomic clock via GNSS signal.

- Generation of optical spectra with a homogeneous intensity of individual comb lines over the full bandwidth. The typical spectral 'flatness' is of 3 dB which is directly suitable for spectrograph calibration and does not require any additional flattening devices. Additionally, such an instrument will hold several advantages from a perspective of a space-based mission:
    - Telecom-grade components that are of substantial maturity and a great potential for miniaturisation and minimisation of the power consumption.
    - Simple system architecture not requiring any optical alignment.

Figure 6 shows a scheme of the laser calibration source. A CW laser operating at a wavelength of 1560 nm is first electro-optically modulated then optically amplified using a standard erbium–doped fibre amplifier (EDFA). The amplification is necessary to bring the optical intensity to such a level that the nonlinear frequency conversion produces enough light for spectrograph calibration. First, the 1560 nm light is frequency doubled in a periodically poled lithium niobate (PPLN) ridge waveguide to generate a narrow band frequency comb around 780 nm. Next, to access the visible wavelength range, the fundamental at 1560 nm and the second harmonic at 780 nm are combined in another PPLN device and via sum-frequency generation (SFG) transfers the light to the 520 nm band. The final output from the LFC module is a beam combining all of the three wavelength ranges. To provide absolute frequencies, the CW laser of the high repetition rate frequency comb can be either stabilised to a standard low repetition rate self-referenced laser frequency comb, itself referenced to a GNSS signal, or to a rubidium two-photon optical transition.

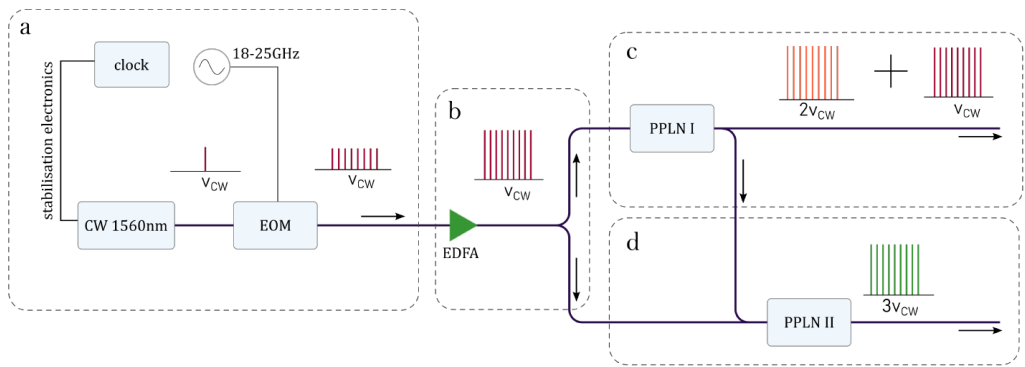

*Figure 6: Scheme of the vANCESTOR's laser system. a) A stabilised continuous wave (CW) laser at 1560 nm is electro-optically modulated (EOM) at a modulation frequency in the range from 18 to 25 GHz. The EOM creates multiple sidebands to the CW line whose separation is equal to the modulation frequency. b) The light is optically amplified in an erbium-doped fibre amplifier (EDFA) and split into two arms. c) In the first arm the light is frequency doubled in a periodically poled lithium niobate (PPLN I) ridge waveguide. The output of this arm is the light at the fundamental frequency $\nu_{CW}$ and at the doubled frequency $2 \cdot \nu_{CW}$. d) Part of the light from the frequency doubling is mixed with the light at the fundamental frequency in a second PPLN to generate the light around the triple value of the fundamental frequencies. The final output of the system consists in three wavelength bands around 1560 nm, 780 nm, and 520 nm combined into one beam.*

## 4.3 GNSS receiver

It is planned to equip the satellite with a GNSS receiver that will provide a frequency reference for the calibration source as well as a very precise determination of the orbit. The precise orbit determination is essential. Indeed, the varying Doppler effect of the satellite viewed by the ground-based spectrograph needs to be compensated for at a high-level of fidelity (ideally to less than 1 cm/s level, radial component of the velocity vector). Equally important is the availability of a GNSS-grade disciplined frequency reference signal in order to reference the broadcasted light back to atomic clocks (GNSS system time). These two key functions, commonly called Position, Navigation and Timing (PNT) will be ensured by a state-of-the-art multi-frequency and multi-constellation GNSS receiver.

### 4.4 Telescope and pointing system

The concept for the optical link between the satellite and the ground telescopes needs to be developed. Optical terminals must adapt to the beam expected by the ground telescopes, such that the satellite is perceived as a star from the ground. The ground-based telescope needs to follow the artificial star for the duration of the wavelength calibration, and as such needs to know the orbit to plan the pointing sequence. Analogously, the space terminal needs to lock then continuously track the ground station. Techniques such as an active pointing correction using e.g. a dedicated ground beacon will be investigated. The pointing will rely on the information provided by the GNSS receiver and the chosen orbit.

The telescope will be designed so that the light beam, on the ground, is not wider than approximately one hundred meters, to maximize the illumination of the pupil of the telescope on the ground, on the one hand, while providing some margin to facilitate pointing. A wider footprint would ease pointing but would require a more intense light source at satellite level to reach a given calibration accuracy. To give an estimate for the telescope size (primary mirror diameter), if we assume that the optics is diffraction limited and the light source is a point (single mode fiber), then a 15 cm telescope will generate a beam including 80% of the energy in a disc of 120 m, if the satellite altitude is 20'000 km and the wavelength 500 nm. If on-going investigations will prove that the power required for the laser is not a design driver, the above footprint could be extended beyond 4" to gain independence from atmospheric influence due to density and temperature gradients, and winds, leading to both elevation dependent refraction and wavelength dependent differential refraction. The selected mirror diameter will not be very different from the value above. Therefore, design and manufacturing of the beam transmission optics will not be a significant challenge in this project: this is considered classical, space optics engineering.

### 4.5 Mission profile

The LFC will be placed on a small 3-axis stabilized Earth observation platform (PF) in a medium Earth orbit (MEO). The orbit is being optimized to have sufficient ground contacts (e.g. any other day during nighttime) for each telescope while maintaining a low angular rate on the sky that can be tracked by standard telescopes. The PF being investigated shall be compatible with a small launcher solution or launched as an auxiliary passenger (Falcon 9, Vega, PSLV).

An ongoing trade-off shall establish if either the PF or the instrument has to be actively pointed towards the ground telescope during the calibration. The required absolute pointing accuracy is of the order 0.5 arcseconds for a 100m diameter footprint.

Calibration would be provided as a service to ground based telescopes; telescopes can register for certain calibration slots. Planning will be done in cycles where all telescope visits planned for the next cycle are registered and the command sequence uplinked to the SC. The SC will then autonomously point the calibration source to the respective ground locations at the indicated times. No real-time interaction between space segment and telescopes is being currently considered.

Detailed knowledge of the relative velocities of PF and telescope is required in order to compensate for the Doppler shift. Currently two options are being studied: a) on-board compensation of the expected doppler frequency shift so that the telescope observes the reference frequency and b) no compensation but provision of detailed time-dependent frequencies at the telescope for post-processing. If option b) is a viable option will depend on the maximum allowable velocity gradient during a calibration exposure and hence on the exposure time of the calibrations.

## 5 TECHNICAL CHALLENGES

A number of enabling technical challenges have been already identified and used to plan within the ongoing study phase a few prototyping activities which main goal is to derisk as much as possible the following development phases (being also compatible with the relatively short time and funds available, see §9).

A likely key requirement for providing absolute optical reference to the Earth-based telescopes is the compensation of the varying Doppler effect which will be achieved by a controlled frequency sweep of the comb lines. In the breadboarding activity, such a high-repetition rate frequency comb will be manufactured and extensively tested. The key objectives are to demonstrate the light generation at the different spectral bands via nonlinear frequency conversion, the ultraprecise frequency sweep to compensate the Doppler shift and its referencing, via a laboratory low-repetition rate frequency comb, to the GNSS signal provided by the GNSS receiver to be breadboarded in this activity. Standard off-the-shelf space-qualified electro-optic modulators are available and will be initially used for breadboarding activities of the high-repetition rate frequency comb, but important reduction of the SWaP can be achieved by transitioning to PIC-based modulators. It is envisaged to develop such PIC-based modulators during the on-going study phase and to implement them, in replacement of the commercially available ones, in the high-repetition rate frequency comb breadboard. By completion of this activity, a consolidated assessment of this promising technology will be available. A fine tuning of the LFC wavelengths bands is ongoing to possibly maximize the usage of νANCESTOR by all major observatories with EPRV spectrographs.

As the calibration precision must be kept at a cm/s level, precise satellite tracking is a critical part that requires hardware development and testing. The same applies for the frequency reference generation function that will be provided and disciplined to GNSS system time by the GNSS receiver. It is envisaged as part of the on-going study phase that a breadboard of an upgraded commercially available GNSS receiver will be developed and extensively tested with respect to positioning and timing / reference frequency disciplining. To do so, the timing (hence the frequency reference) and the on-board algorithms for orbit determination and timing / reference frequency discipling will be further improved. The MEO of the mission means that, depending on the final orbit configuration, observation of GNSS sidelobe signals and GNSS antennas in both zenith and nadir directions might be necessary. As of today, the commercial performance of $10^{-10}$ fractional instability at an averaging time of 1 second is sufficient to meet usual commercial targets. However, those figures are insufficient for this mission. An RF upgrade will be implemented with limited hardware modifications, with the plan to obtain a one to two order of magnitude improvement, i.e. a fractional instability of $10^{-11} – 10^{-12}$ at an averaging time of 1 second.

For successful pointing of the calibration laser beam, the different error sources must be smaller than the achievable accuracy. This imposes limits on the static and slowly varying deterministic errors and on the random and/or rapidly oscillating errors, which cannot be compensated. If chromatic pointing differences exist, they will eventually have to be reduced by differential point-ahead paths. The systematic pointing errors being in part unknown, a pointing calibration method shall be investigated.

## 6 PROJECT STATUS

The νANCESTOR project has been granted funding within Switzerland to perform a phase 0 and phase A study lasting 36 months in total. The project has been officially kicked off on 2024 April 1st, with a target completion date by 2027 March 31. The initial phase 0 is being exploited in order to prepare the requirement documents of the main mission components (the LFC, the on-board telescope and the GNSS receiver), as well as the science requirement document and the requirement document on the orbit. The phase 0 will be completed by the end of 2024, holding a formal Mission Requirements Review (MRR) and allowing the νANCESTOR Swiss consortium to release a call for the participation of the Swiss industries to prototyping activities planned for the phase A (and based on the set of requirement documents completed in phase 0). The phase A will be formally closed after the successful closure of a Mission Definition Review (MDR), which include the evaluation of a number of prototyping activities and associated tests on the most critical technological aspects (see details in the preceding sections). Both the MRR and the MDR will be formally led by the ESA PRODEX office with the support of the relevant ESA technical departments. Several Swiss institutions are currently supporting the on-going phase 0 study. These include the University of Geneva, providing the overall project management and the scientific lead, the University of Bern, providing the required knowledge in the field of mission definition and platform design, the ETH Zurich, filling in the experience in the field of GNSS devices, and the HES-SO, mainly involved in the definition of the on-board telescope and the links between the platform and the ground telescopes. As mentioned earlier in the paper, the consortium is relying on CSEM for all knowledge and plans on the LFC and system engineering.

## ACKNOWLEDGEMENTS

The Swiss consortium acknowledges support from the Swiss Space Office for the on-going study phase and support from ESA PRODEX for most of the technical aspects of νANCESTOR under active and intense investigation.